\documentstyle[prb,aps,twocolumn,epsfig]{revtex}

\title{Free-electron Model for Mesoscopic Force Fluctuations in Nanowires}  

\author{S.~Blom$^{1}$, H.~Olin$^{2}$, J.~L.~Costa-Kr{\"a}mer$^{3}$, N.~Garc{\'\i}a$^{3}$, M. Jonson$^{1}$, P.~A.~Serena$^{3}$ and R.~I.~Shekhter$^{1}$}

\address{
 $^{1}$Department of Applied Physics and $^{2}$Department of Physics,\\ 
Chalmers University of Technology and G\"oteborg University, 
S-412 96 G\"oteborg, Sweden\\
 $^{3}$Laboratorio de Fisica de Sistemas Peque{\~n}os y 
Nanotecnologia,\\
Consejo Superior de Investigaciones Cientificas (CSIC), Serrano 144, E-28006 Madrid, Spain \\
}

\begin{document}

\maketitle

\begin{abstract}
When two metal electrodes are separated, a nanometer sized wire (nanowire) is formed just before the contact breaks. The electrical conduction measured during this retraction process shows signs of quantized conductance in units of $G_0=2e^2/h$. Recent experiments show that the force acting on the wire during separation fluctuates, which has been interpreted as being due to atomic rearrangements. In this report we use a simple free electron model, for two simple geometries, and show that the electronic contribution to the force fluctuations is comparable to the experimentally found values, about 2 nN.
\end{abstract}
\section{Introduction}
The electrical conductance through a narrow constriction with a diameter of the order of the electron wavelength is quantized in units of $G_0=2e^2/h$ \cite{Landauer89,Beenakker91}. Such conductance quantization is observed at low temperatures  in semiconductor devices containing a two-dimensional electron gas (2DEG) \cite{Wharam88,vanWees88}. Similar effects are possible\cite{Garcia87,Garcia89} at room temperature in metallic wires with a diameter of the order of 1 nm (nanowires) and are observed using scanning tunnelling microscopy (STM)\cite{Pascual93,Pascual95,Olesen94,Brandbyge95,Agrait93,C-K97a,C-K97b}, 
mechanically controlled break junctions (MCBJ)\cite{Muller92,Krans95} or, as recently shown\cite{C-K95,Garcia96}, just by using plain macroscopic wires. These techniques use the same basic principle: by pressing two metal pieces together a metallic contact is formed which can be stretched to a nanowire by the subsequent separation of the electrodes. The conductance in such a system is found to decrease in abrupt steps with a height of about $2e^2/h$, just before the contact breaks.

In a recent pioneering experiment by Rubio, Agrait, and Vieira \cite{Rubio96}, following earlier attempts\cite{Agrait95,Stalder96a,Landman90,Stalder96b}
, the force and the conductance were simultaneously measured during elongation, from formation to rupture, of a gold nanowire. They show that the stepwise variation of the conductance is always accompanied by an abrupt change  of the force. One interpretation\cite{Brandbyge95,Rubio96,Landman96,Torres96,Todorov96} is that the structural transformations of the nanowire, involving elastic and yielding stages, cause the stepwise variation of the conductance.

In this report we study the electronic contribution to the observed force fluctuations using a simple free electron approach neglecting all atomic structures of the wire: a jellium model (see also three other recent reports \cite{Stafford97,vanRuitenbeek97,Yannouleas97}). In metals the electronic contribution to the binding energy is significant (metallic binding) and one might suspect that the quantized electronic energy levels in the nanowire would be reflected in the binding energy. When a conductance mode closes it should produce a sharp change in the electronic binding energy, and subsequently the force. The quantized energy levels are of the order of eV and the wire elongation of the order of nm giving a change in force of the order of  nN, the same as observed in the experiments. Considering this, we develop in this report a simple free electron model. The calculations show force fluctuations of the same size as in the experiments.

\begin{figure}[htb]
\vspace{0.5cm}
\epsfig{file=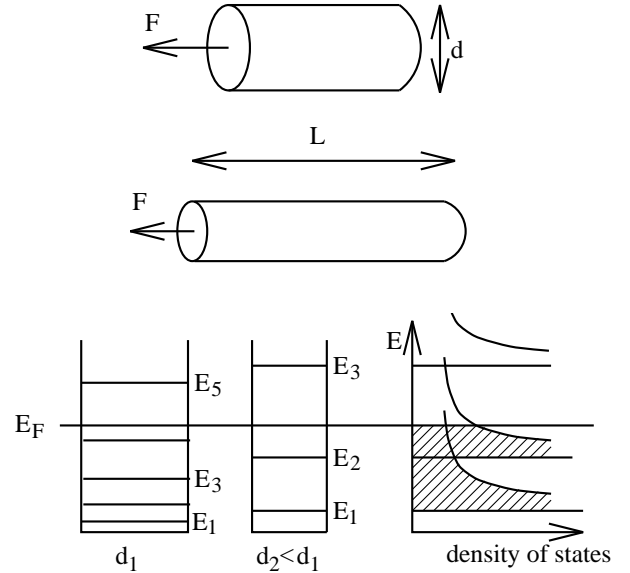, width=7.7cm}
\vspace{0.5cm}
\caption{Model of the nanowire. The pulling force $F$ is acting on a nanowire of length $L$ and width $d$. When the wire is elongated more and more transverse quantized modes are pushed above the Fermi level and closed. The corresponding one-dimensional densities of states are filled up to the Fermi level.}
\label{fig:model}
\end{figure}

\section{Model}

We use a free electron model neglecting all atomic structure in the wire, a jellium model. Further,  cylindrical nanowires of length $L$ and with two different cross-sections are studied: first with a circular cross-section, see figure~\ref{fig:model}, and then with a square cross-section. Under the assumed ideal plastic deformation, the volume $V$ of the wire will be constant during elongation. We are interested in the tensile forces acting on the wire during elongation. In general the force is the derivative of the energy with respect to distance, however, our system is open and we should instead consider the thermodynamic potential, $\Omega$. The Fermi energy, $E_F$, in metals is much higher than the thermal energy and we can approximate the chemical potential by $E_F$ and the thermodynamic potential is found to be $\Omega=E-E_FN$, where E is the energy and N the number of electrons. The force is then $F=-d\Omega/dL$.

\subsection{Nanowire with circular cross-section}\label{ss:circ}
In a wire with circular cross-section, using the adiabatic approximation\cite{Glazman88}, the  transverse motion of the electrons gives rise to quantized modes $n$ of energy, 
\begin{equation}
E_n^c = E_{jl} = \frac{\hbar^2}{2mR^2}\beta_{jl}^2=\frac{\hbar^2\pi}{2mV}L\beta_{jl}^2 \label{eq:bottom}
\end{equation}
where $\beta_{jl}$ are roots to Besselfunctions, i.e. $\beta_{jl}=2.4048,3.8317,\ldots $, and the degeneracy is two-fold (not counting spin) unless $j$ equals zero. The third equality in equation~\ref{eq:bottom} is valid for a wire of constant volume. A mode is considered to be open if $E_F > E_n^c$. The number of electrons in the wire is
\begin{eqnarray}
N & = & \sum_nN_n=\sum_n\int_{E_n^c}^{E_F}LD(\varepsilon-E_n^c)d\varepsilon\\
  & = &\sum_n2L\sqrt{\frac{2m}{\pi^2\hbar^2}}\sqrt{E_F-E_n^c}
\end{eqnarray}
where $D(\varepsilon)$  is the one-dimensional density of states:
\begin{equation}
D(\varepsilon)=\sqrt{\frac{2m}{\pi^2\hbar^2}}\frac{1}{\sqrt{\varepsilon}}
\end{equation}
The total electronic energy of the wire is the integral of the energy
times the density of states up to the Fermi energy  and summed over all
open modes,
\begin{eqnarray}
E & = & \sum_n\int_{E_{n}^c}^{E_F}L\varepsilon D(\varepsilon-E_{n}^c)d\varepsilon\\
  & = & \sum_n\left\{\frac{2}{3}L\sqrt{\frac{2m}{\pi^2\hbar^2}}\left(E_F-E_n^c\right)^{3/2}+E_n^cN_n\right\}
\end{eqnarray}
The thermodynamic potential is then 
\begin{equation}
\Omega = E-E_FN=-\sum_n\frac{4}{3}L\sqrt{\frac{2m}{\pi^2\hbar^2}}\left(E_F-E_n^c\right)^{3/2}
\end{equation}
and the derivative gives the force,
\begin{eqnarray}
F = -\frac{d\Omega}{dL}=&&\sum_n\sqrt{\frac{2m}{\pi^2\hbar^2}}\left\{\frac{4}{3}\left(E_F-E_n^c\right)^{3/2}\right.\\
&&\left.-2\left(E_F-E_n^c\right)^{1/2}E_n^c\right\} \label{eq:force}
\end{eqnarray}

\subsection{Nanowire with square cross-section}
Using the same approach as above, the  transverse motion of the electrons in the wire with square cross-section, gives rise to quantized modes $n$ of energy,
\begin{equation}
E_n^q = \frac{\hbar^2\pi^2}{2md^2}n^2=\frac{\hbar^2\pi^2}{2mV}n^2L \label{eq:enq}
\end{equation}
where $n^2=l^2+m^2$, $l=1,2,\ldots$ and $m=1,2,\ldots$, i.e. $n^2=2,5,8,10,\ldots$ and the degeneracy is two-fold (not counting spin) unless $l$ and $m$ are equal. The second equality in equation~\ref{eq:enq} is valid for a wire of constant volume. Replacing $E_n^c$ with $E_n^q$ in subsection~\ref{ss:circ} gives the appropriate expressions for the wire with square cross-section.

\section{Force Fluctuations} 

Figure~\ref{fig:results} shows a plot of the force during elongation, in a wire with circular cross-section, according to equation~\ref{eq:force}. Also the conductance of the wire is shown. The wire-volume is taken to be 3 nm$^3$. The number of modes that contribute to the conductance is taken from equation~\ref{eq:bottom}. Whenever a mode closes the conductance jumps one quantum unit and an abrupt change in the force appears. The peak-to-peak amplitude of the force fluctuations between two modes is about 2 nN. 

Figure~\ref{fig:results_sq} shows the corresponding force and conductance for a wire with square cross-section.

\section{Discussion}
\begin{figure}
\begin{centering}
\epsfig{file=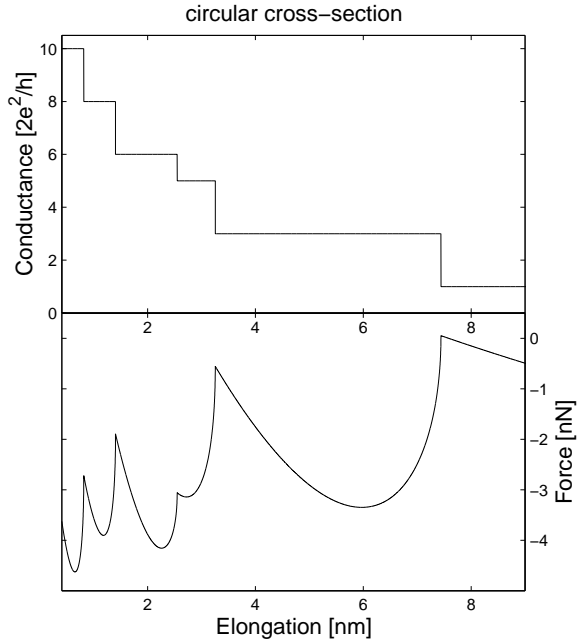, width=7.7cm}
\caption{Calculated conductance and force as a function of elongation
for gold nanowires with circular cross-section and constant volume 3 nm$^3$. Whenever a mode closes the conductance jumps one quantum unit and an abrupt change in the force appears. The peak-to-peak amplitude of the force fluctuations between two modes is about 2 nN. }
\label{fig:results}
\end{centering}
\end{figure}

\begin{figure}
\begin{centering}
\epsfig{file=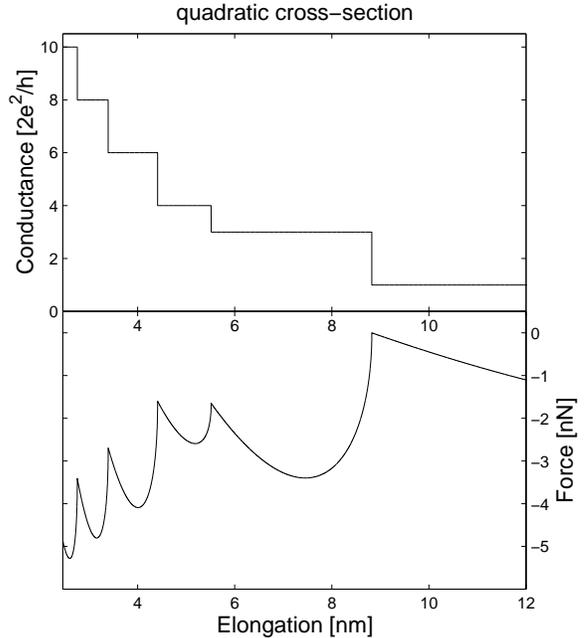, width=7.7cm}
\caption{Calculated conductance and force as a function of elongation
for gold nanowires with square cross-section and constant volume  3 nm$^3$.}
\label{fig:results_sq}
\end{centering}
\end{figure}

The force from our calculations shown in figure~\ref{fig:results} and figure~\ref{fig:results_sq} agree both qualitatively as well as quantitatively with experiments \cite{Rubio96}. The only significant effect of the geometry of the cross-section is on the degeneracy of the modes.

Force fluctuations are also seen in molecular dynamics simulations \cite{Brandbyge95,Landman90,Landman96,Todorov96} and the jumps in conductance are interpreted as due to atomic rearrangements. However, because of the experimental like conditions in these simulations, it is difficult to separate the different contributions to the binding energy. Our interpretation is more or less the reversed: the electronic contribution to the binding energy is so large that the change of the quantized energy levels in the wire, with a corresponding quantized conductance, causes the force fluctuations. These force fluctuations might then give rise to atomic rearrangements but not necessarily. Although this is a bit like the story about the hen and the egg, our simple model shows that the electronic contribution must be considered seriously because it constitutes a significant part of the metallic binding energy in these nanowires.

One electronic contribution to the binding energy which is neglected in the present model is the Coulomb interaction. In metallic binding the electrostatic energy could be of the same order as the kinetic one and a natural extension of the present model would be to include the electrostatic energy which would change the electronic energy. The force fluctuations would, however, still be present.

\section{Conclusion}
We have shown, using a simple free electron model, that the electronic contribution to the force fluctuations is comparable to the experimentally found values. This could be of importance to understand the mechanism of formation of metallic nanowires as well as in the wider context of nanomechanics.

\section*{Acknowledgement}
This work was supported by the European ESPRIT project Nanowires, the Spanish DGCIT and CICyT, and the Swedish NFR and TFR agencies.


\begin{thebibliography}{99}

\bibitem{Landauer89}
R. Landauer, J. Phys. Condens. Matter {\bf 1}, 8099 (1989).

\bibitem{Beenakker91}
C. W. J. Beenakker and H. van Houten, in {\it Solid State Physics}, Vol. 44,(1991).

\bibitem{Wharam88}
D.~A.~Wharam, T.~J.~Thornton, R.~Newbury, M.~Pepper, H.~Ahmed, J.~E.~F.~Frost, D.~G.~Hasko, D.~C.~Peacock, D.~A.~Ritchie and G.~A.~C.~Jones, J. Phys. C {\bf 21}, L209 (1988).

\bibitem{vanWees88}
B.~J.~van~Wees, H.~van~Houten, C.~W.~J.~Beenakker, J.~G.~Williamson, D.~van~der~Marel and C.~T.~Foxton, Phys. Rev. Lett. {\bf 60}, 848 (1988).

\bibitem{Garcia87}
N.~Garcia, STM Workshop, ICTP Trieste 1987, (unpublished).

\bibitem{Garcia89}
N.~Garcia and L.~Escapa, Appl. Phys. Lett. {\bf 54}, 1418 (1989).

\bibitem{Pascual93}
J.~I.~Pascual, J.~Mendez, J.~G\'omez-Herrero, A.~M.~Baro, N.~Garcia and V.~T.~Binh, Phys. Rev. Lett. {\bf 71}, 1852 (1993).

\bibitem{Pascual95}
J.~I.~Pascual, J.~Mendez, J.~G\'omez-Herrero, A.~M.~Baro, N.~Garcia, U.~Landman, W.~D.~Luedtke, E.~N.~Bogachek and H.~P.~Cheng, Science {\bf 267}, 1793 (1995).

\bibitem{Olesen94}
L.~Olesen, E.~L\ae gsgaard, I.~Stensgaard, F.~Besenbacher, J.~Schi\o tz, P.~Stoltze, K.~W.~Jacobsen and J.~K.~N\o rskov, Phys. Rev. Lett. {\bf 72}, 2251 (1994).

\bibitem{Brandbyge95}
M.~Brandbyge, J.~Schi\o tz, M.~R.~S\o rensen, P.~Stoltze, K.~W.~Jacobsen, J.~K.~N\o rskov, L.~Olesen, E.~L\ae gsgaard, I.~Stensgaard and F.~Besenbacher, Phys. Rev. B {\bf 52}, 8499 (1995).

\bibitem{Agrait93}
N.~Agrait, J.~G.~Rodrigo and S.~Vieira, Phys. Rev. B {\bf 47}, 12345 (1993).

\bibitem{C-K97a}
J.~L.~Costa-Kr\"amer, N.~Garcia and H.~Olin, Phys. Rev. B {\bf 55}, 12910 (1997).

\bibitem{C-K97b}
J.~L.~Costa-Kr\"amer, N.~Garcia and H.~Olin, Phys. Rev. Lett. {\bf 78}, 4990 (1997).

\bibitem{Muller92}
C.~J.~Muller, J.~M.~van~Ruitenbeek and L.~J.~de~Jongh, Phys. Rev. Lett. {\bf 69}, 140 (1992).

\bibitem{Krans95}
J.~M.~Krans, J.~M.~van~Ruitenbeek, V.~V.~Fisun, I.~K.~Yanson and L.~J.~de~Jongh, Nature {\bf 375}, 767 (1995).

\bibitem{C-K95}
J.~L.~Costa-Kr\"amer, N.~Garcia, P.~Garcia-Mochales and P.~A.~Serena, Surf. Sci. Lett. {\bf 342}, L1144 (1995).

\bibitem{Garcia96}
N.~Garcia and J.~L.~Costa-Kr\"amer, Europhysics News {\bf 27}, 89 (1996).

\bibitem{Rubio96}
G.~Rubio, Agrait and S.~Vieira, Phys. Rev. Lett. {\bf 76}, 2302 (1996).

\bibitem{Agrait95}
N.~Agrait, G.~Rubio and S.~Vieira, Phys. Rev. Lett. {\bf 74}, 3995 (1995).

\bibitem{Stalder96a}
A.~Stalder and U.~D\"urig, J. Vac. Sci. Techn. B {\bf 14}, 1259 (1996).

\bibitem{Landman90}
U.~Landman, W.~D.~Luedtke, N.~A.~Burnham and R.~J.~Colton, Science {\bf 248}, 454 (1990).

\bibitem{Stalder96b}
A.~Stalder and U.~D\"urig, Appl. Phys. Lett. {\bf 68}, 637 (1996).

\bibitem{Landman96}
U.~Landman, W.~D.~Luedtke, B.~E.~Salisbury and R.~L.~Whetten, Phys. Rev. Lett. {\bf 77}, 1362 (1996).

\bibitem{Torres96}
J.~A.~Torres and J.~J.~Saenz, Phys. Rev. Lett. {\bf 77}, 2245 (1996).

\bibitem{Todorov96}
T.~N.~Todorov and A.~P.~Sutton, Phys. Rev. B {\bf 54}, 14234 (1996).

\bibitem{Stafford97}
C.~A.~Stafford, D.~Baeriswyl and J.~B\"urki,  Phys. Rev. Lett. {\bf 79}, 
2863 (1997). 

\bibitem{vanRuitenbeek97}
J.~M.~van~Ruitenbeek, M.~H.~Devoret, D.~Esteve and C.~Urbina, (unpublished).

\bibitem{Yannouleas97}
C.~Yannouleas and U.~Landman, J. Phys. Chem. A {\bf 101}, 4780 (1997).

\bibitem{Glazman88}
L.~I.~Glazman, G.~B.~Lesovik, D.~E.~Khmel'nitskii and R.~I.~Shekhter, JETP Lett. {\bf 48}, 238 (1988).

\bibitem{Bogachek90}
E.~N.~Bogachek, A.~M.~Zagoskin and I.~O.~Kulik, Fiz. Nizk. Temp. {\bf 16}, 1404 (1990) [Sov. J. Low Temp. Phys. {\bf 16}, 796 (1990)].

\end{thebibliography}
\end{document}